\documentstyle[preprint,eqsecnum,aps,psfig]{revtex}
\begin{document}
\draft
\title{$Z_2$-Regge versus Standard Regge Calculus in two dimensions}
\author{E.~Bittner, A.~Hauke, H.~Markum, J.~Riedler
}
\address{Institut f\"{u}r Kernphysik, Technische Universit\"{a}t Wien,
 A-1040 Vienna, Austria}
\author{C.~Holm}
\address{Max-Planck-Institut f\"ur Polymerforschung, D-55128 Mainz, Germany}
\author{W.~Janke}
\address{Institut f\"ur Physik, Johannes Gutenberg-Universit\"at Mainz,
 D-55099 Mainz, Germany\\
Institut f\"ur Theoretische Physik, Universit\"at Leipzig, D-04109 Leipzig,
Germany}
\date{\today}
\maketitle
\begin{abstract}
We consider two versions of quantum Regge calculus. The Standard Regge
Calculus where the quadratic link lengths of the simplicial manifold vary
continuously and the $Z_2$-Regge Model where they are restricted to two
possible values. The goal is to determine whether the computationally more
easily accessible $Z_2$ model still retains the universal characteristics
of standard Regge theory in two dimensions. In order to compare observables
such as average curvature or Liouville field susceptibility, we use in both
models the same functional integration measure, which is chosen to render
the $Z_2$-Regge Model particularly simple. Expectation values are computed
numerically and agree qualitatively for positive bare couplings. The phase
transition within the $Z_2$-Regge Model is analyzed by mean-field theory.
\end{abstract}

\narrowtext

\section{Introduction} \label{intro}
Standard Regge Calculus (SRC) \cite{regge} provides an interesting method to
explore quantum gravity in a non-perturbative fashion \cite{regge_background}.
The infinite degrees of freedom of Riemannian manifolds are reduced by
discretization, that is, SRC deals with piecewise linear spaces described
by a finite number of parameters. A manifold is approximated by a simplicial
lattice with {\em fixed} coordination numbers, as opposed to the
dynamical triangulated random surface (DTRS) method \cite{kazakov}
where the coordination numbers are treated as the dynamical degrees of freedom.
This leaves in SRC the quadratic link lengths
$q$ as gravitational degrees of freedom which are constraint by triangle
inequalities. Since analytical treatments have
proven to be difficult, this approach has been extensively studied
through numerical computer simulations during the last ten years
\cite{nonpert,ham}. Although the computer codes can be
efficiently vectorized, large scale simulations are still a very time
demanding enterprise. One therefore seeks for suitable approximations
which will simplify the SRC and yet retain most of its universal
features.

The $Z_2$-Regge Model ($Z_2$RM) \cite{z2rm} could be such a
desired simplification. Here the quadratic link lengths $q$ of the simplicial
complexes are restricted to take on only the two values
$q_l=1+\epsilon\sigma_l, 0<\epsilon <\epsilon_{max}, \sigma_l =\pm 1$,
in close analogy to the ancestor of all lattice models, the Lenz-Ising model.
To test whether this simpler model is in a reasonable sense still similar to
SRC, i.e. shares the same universal properties, we study both models in two
dimensions and compare a number of
observables for one particular lattice size. Moreover we estimate in both
models the critical exponent $\eta_\phi$ of the Liouville field susceptibility
by performing a finite-size scaling analysis on moderately sized lattices.

Although some models for 2d-quantum gravity have been exactly solved via
the matrix model approach \cite{matrixmodel} and with the help of conformal
field theory \cite{kpz}, the relation of those approaches to SRC is not yet
understood. There are even severe discrepancies between the alternative
discrete approach, the DTRS method \cite{kazakov}, and SRC.
Especially the functional integration measure in SRC is under
heavy debate \cite{amham}.
In an effort to clarify the role of the measure the conventional
definition of diffeomorphisms has been employed,
assuming that a piecewise linear space, i.e. a Regge surface, is
{\it exactly} invariant under the action of the full diffeomorphism
group \cite{meno}. After a conformal gauge fixing was performed in
the continuum formalism, it was shown that the evaluation of the
non-local Faddeev-Popov determinant by using such a Regge regularization
leads to the usual Liouville field theory results in the continuum
limit. All that is based on a description of piecewise linear manifolds
with deficit
angles, not edge lengths, and is mostly taken as an argument that
the correct measure of Standard Regge Calculus has to be non-local. However,
to our knowledge
it is not obvious that this argument carries over
to a discretized Lagrangian, which is formulated in terms of fluctuating
edge lengths, obeying triangle equalities, and which is {\em not} invariant
under the
diffeomorphism group due to the presence of curvature defects: different
assignments of edge lengths correspond to
different physical geometries, and as a consequence there are no gauge degrees
of freedom in Standard Regge Calculus, apart from special geometries
(like flat space) \cite{hartle}. Therefore we do not include a gauge fixing
term and rely in this work still on a local measure. Our
main goal in this present investigation is not to resolve the measure question
but to explore the phase behavior of the $Z_2$RM and its relation to SRC.
We will show that the discretized $Z_2$RM does not suffer from
unphysical gauge degrees of freedom. If both SRC and $Z_2$RM
for certain local measures lie in the same
universality class one can hope to learn about physical observables
using this simplified approach.

The rest of the paper is organized as follows: In Chapter \ref{sect2} we
briefly review the Standard Regge Model as well as the $Z_2$-Regge Model.
In Chapter \ref{sect3} we introduce the observables and discuss important
scaling relations. The details of the Monte Carlo simulations and the
results are presented in Chapter \ref{sect4}. Chapter \ref{sect5} deals
with a mean-field approach for the $Z_2$RM to discuss the observed phase
transition in that specific model. Finally Chapter \ref{sect6} ends with
our conclusion.

\section{Models} \label{sect2}
Starting point for both Standard Regge Calculus and the $Z_2$-Regge
Model is Regge's discrete description of General Relativity in
which space-time is represented by a piecewise flat, simplicial manifold:
the Regge skeleton \cite{regge,nonpert}. The beauty of this procedure is
that it works for any space-time dimension $d$ and for metrics of arbitrary
signature. The Einstein-Hilbert action translates into
\begin{equation} \label{Iq}
I(q) = \lambda\,\sum_{s^d}\,V(s^d) - 2\beta\,\sum_{s^{d-2}}\delta(s^{d-2})
V(s^{d-2})~,
\end{equation}
with the quadratic edge lengths $q$ describing the dynamics of the lattice,
$\lambda$ being the cosmological constant, and $\beta$ the bare Planck mass
squared.
The first sum runs over all $d$-simplices $s^d$ of the simplicial complex and
$V(s^d)$ is the $d$-volume of the indicated simplex. The second term represents
the curvature of the lattice, that is concentrated on the $(d-2)$-simplices
leading to deficit angles $\delta(s^{d-2})$, and is proportional to the
integral over the curvature scalar in the classical Einstein-Hilbert action
of the continuum theory. The connectivity of the edges, in simplicial
terminology called the incidence matrix, is fixed from the beginning through
the simplicial decomposition of the manifold under consideration.
Any smooth manifold can be approximated by a Regge skeleton with
arbitrarily small deficits simply by using a sufficient number of links and
arranging them appropriately.

\subsection{Standard Regge Calculus}
In two dimensions Regge's discretization procedure is easily illustrated by
choosing a triangulation of the considered surface. Each triangle
then represents a part of a piecewise linear manifold. The net of triangles
itself is a two-geometry, with singular (non-differentiable) points located
at the vertices of the net. In the presence of curvature a vector that is
parallel transported around a vertex experiences a rotation by the
deficit angle $\delta _i=2\pi -\sum_{t\supset i}{\theta_i(t)}$, where
$\theta_i(t)$ are the dihedral angles of the triangles $t$ attached to
vertex $i$.
The integral of the scalar curvature over the simplicial complex $K$ in
two dimensions is a topological invariant due to the Gauss-Bonnet theorem.
Its simplicial analogue reads as
\begin{equation} \label{GB}
\sum_{i\supset K} \delta_i = 2\pi\chi({\cal M})~,
\end{equation}
where $\chi=2(1-g)$ is the Euler characteristic of the manifold
${\cal M}$ expressed
by the number of handles $g$ in ${\cal M}$.

In the exceptional case of flat skeletons one can move a vertex on the
surface, keeping all the neighbors fixed, without violating the triangle
inequalities, such that different configurations triangulate the same (flat)
geometry. This transformation has two parameters and is an exact invariance of
the action, but does not exist in general. When space is curved the invariance
is only an approximate one. In the limit of increasing number of links local
gauge invariance, that is the continuum diffeomorphism group should be
recovered.

A quantization of the above action (\ref{Iq}) proceeds by evaluating the
path integral
\begin{equation} \label{Z2RC}
Z=\int\!D[q] e^{-I(q)}~.
\end{equation}
In principle the functional integration should extend over all metrics on
all possible topologies, but, as is usually done, we restrict ourselves to
one specific topology, the torus, ${\cal M}=T^2$. Consequently the Euler
characteristic $\chi(T^2)$ vanishes in (\ref{GB}) and the action (\ref{Iq})
consists only of a cosmological constant $\lambda$ times the sum over all
triangle areas $A_t$. The path-integral approach suffers from a non-uniqueness
of the integration measure, even the need for a non-local measure is
advocated. However, some of the proposed
non-local measures do not agree with their continuum counterparts in the weak
field limit, which is a necessary condition for an acceptable discrete
measure \cite{amham}. This property however is fulfilled
for the standard simplicial measure \cite{ac}
\begin{equation} \label{simeas}
\int\!D[q] = \Pi_l\,\int \frac{d q_l}{q_l^m} {\cal F}_\eta({q_l})~,
\end{equation}
with $m\in I\!\!R$ permitting to investigate a 1-parameter family of measures.
The function ${\cal F}_\eta(q_l)$ constrains the integration to those
Euclidean configurations of link lengths which do not violate the triangle
inequalities. The positive parameter $\eta$ modifies the triangle inequalities
to $l_3 \le (l_1 + l_2)(1-\eta)$ and $l_3 \ge |l_1 - l_2|(1 + \eta)$, so that
very thin triangles are suppressed. This is not necessary on theoretical
grounds, but will be useful for the Monte Carlo evaluation of the path
integral.

Hence the model considered here is characterized by the partition function
\begin{equation} \label{CRM}
Z = \left[ \prod _l ^{N_1} \int_0^\infty
dq_lq_l^{-m} \right] {\cal F}_\eta (\{q_l\}) e^{-\lambda\sum _i A_i}~,
\end{equation}
where $N_1$ is the number of links and $A_i = \sum_{t \supset i} A_t/3$
denotes the barycentric area with $A_t$ being the area of a triangle $t$.
A specific choice of the value of the parameter $m$ will be discussed in
the next section.

As discussed above,
the numerical computations of (\ref{CRM}) (for technical details see
Chapter \ref{sect4}) do not run into
the diffeomorphism problem by summing over distinct simplicial lattices
without fixing a gauge. Still the question arises whether one double-counts
some classes of geometries in this way, e.g., one may argue from the above that
flat geometries are over-represented, though simulations give no hints on
that.

\subsection{${\bf Z_2}$-Regge Model}
In the $Z_2$-Regge Model \cite{z2rm} the squared edge lengths $q_l$
are allowed to take on only the two values
\begin{equation} \label{q}
q_l=1+\epsilon\sigma_l ~,\quad
0\le\epsilon <\epsilon_{max} ~, \quad \sigma_l = \pm 1 ~,
\end{equation}
where the parameter $\epsilon$ is chosen such that the Euclidean triangle
inequalities are fulfilled for all $q_l$'s, i.e. ${\cal F}_\eta=1$ for all
configurations $\{q_l\}$.
There exist $2^{N_1}$ different configurations and for finite $\epsilon$
and link lengths none of them can be transformed smoothly into each other.
A further nice attribute of the $Z_2$RM is its accordance with lattice
perturbation theory. As described in \cite{JeNi}, (\ref{q}) can be viewed
as weak-field expansion around flat space, implicitly having performed a
conformal gauge fixing. On each triangle the metric tensor assumes the
form $g_{\mu\nu}(\triangle)=(1+\epsilon)\delta_{\mu\nu}$.
However a finite $\epsilon$ inhibits local conformal transformations. Since
triangles share links, rescaling lengths on one particular triangle would
necessitate the same rescaling for the neighboring triangle and so on. What
remains only is a global transformation. In the continuum limit the local
conformal degree of freedom should be restored.

Using (\ref{q}) the measure (\ref{simeas}) can be replaced by
\begin{equation} \label{pro}
\sum\limits_{\sigma_l = \pm 1}\exp[-m\sum_l\ln(1+\epsilon\sigma_l)] =
\sum\limits_{\sigma_l = \pm 1}\exp[-N_1m_0(\epsilon) -
\sum_l m_1(\epsilon)\sigma_l]~,
\end{equation}
where $m_0 = -\frac{1}{2}m\epsilon^2 + O(\epsilon^4)$ and
$m_1 = m[\epsilon + \frac{1}{3}\epsilon^3 + O(\epsilon^5)]=mM$ with
\begin{equation}
M=\sum_{i=1}^{\infty}\frac{\epsilon^{2i-1}}{2i-1}=\frac{1}{2}\ln
\frac{1+\epsilon}{1-\epsilon}~.
\label{eq:M}
\end{equation}

The area of a single triangle $t$ with squared edge lengths $q_1, q_2, q_l$
can be expressed as
\begin{eqnarray} \label{A}
A_t &=&\frac{1}{2}\left| \begin{array}{cc}
       q_1 & \frac{1}{2}(q_1+q_2-q_l) \\
       \frac{1}{2}(q_1+q_2-q_l) & q_2
       \end{array} \right|^\frac{1}{2} = \nonumber \\[2mm]
    &=&\frac{1}{2}\{\frac{3}{4}+\frac{1}{2}(\sigma_1+\sigma_2+\sigma_l)
       \epsilon + \frac{1}{2}(\sigma_1\sigma_2+\sigma_1\sigma_l+
       \sigma_2\sigma_l-\frac{3}{2})\epsilon^2\}^\frac{1}{2} ~.
\end{eqnarray}
Expanding $A_t$ the series consists only of terms up to $\sigma^3$ since
$\sigma_i^2=1$. Therefore $A_t$ can be written as
\begin{equation} \label{at}
A_t=c_0(\epsilon)+c_1(\epsilon)(\sigma_1+\sigma_2+\sigma_l)+
c_2(\epsilon)(\sigma_1\sigma_2+\sigma_1\sigma_l+\sigma_2\sigma_l)+
c_3(\epsilon)\sigma_1\sigma_2\sigma_l~.
\end{equation}
Computing the four possible values for the triangle areas and comparing with
(\ref{at}) results in exact solutions for the coefficients $c_i$,
\begin{eqnarray} \label{c2}
c_0&=&\frac{1}{32}\left[2\sqrt{3}+3\sqrt{(1-\epsilon)(3+5\epsilon)}
        +3\sqrt{(1+\epsilon)(3-5\epsilon)}\,\right]~, \nonumber\\
c_1&=&\frac{1}{32}\left[2\sqrt{3}\epsilon +\sqrt{(1-\epsilon)
        (3+5\epsilon)}-\sqrt{(1+\epsilon)(3-5\epsilon)}\,\right]~, \nonumber\\
c_2&=&\frac{1}{32}\left[2\sqrt{3}-\sqrt{(1-\epsilon)(3+5\epsilon)}
        -\sqrt{(1+\epsilon)(3-5\epsilon)}\,\right]~, \nonumber\\
c_3&=&\frac{1}{32}\left[2\sqrt{3}\epsilon -3\sqrt{(1-\epsilon)
        (3+5\epsilon)}+3\sqrt{(1+\epsilon)(3-5\epsilon)}\,\right]~.
\end{eqnarray}
Obviously one must have $\epsilon <\frac{3}{5}=\epsilon_{max}$ for the
triangle areas to be real and positive.
In the simulations described below we used $\epsilon = 0.5$
but the special choice of $\epsilon$ does not affect our numerical
results except for the value of the critical coupling ($\lambda_c$)
as one can infer from the next equation. It is obtained by
inserting (\ref{pro}) and (\ref{at}) into the partition function (\ref{CRM})
\begin{equation}
\label{Z}
Z=\sum_{\sigma_l=\pm 1}J\exp\{-\sum_l(2\lambda c_1 + m_1)\sigma_l -
\lambda\sum_t[c_2(\sigma_1\sigma_2 + \sigma_1\sigma_l + \sigma_2\sigma_l)
+ c_3\sigma_1\sigma_2\sigma_l]\}~,
\end{equation}
with $J=\exp(-\lambda N_2c_0-N_1m_0)$ and $N_2$ denoting the total number of
triangles. If we view $\sigma_l$ as a spin variable
and identify the corresponding link $l$ of the triangulation with a
lattice site, then $Z$ reads as the partition function of a
spin system with two- and three-spin
nearest-neighbor interactions on a Kagom{\'e} lattice.
A particularly simple
form of
Eq.~(\ref{Z})
is obtained if one chooses $m_1=-2\lambda c_1$, and therefore
\begin{equation}
m=-\frac{2\lambda c_1}{M} ~,
\label{eq:m_vs_M}
\end{equation}
which is henceforth used for the measure in the $Z_2$RM as well as in SRC.

\section{Observables and Scaling relations} \label{sect3}
To compare both models we examined the quadratic link lengths and the area
fluctuations on the simplicial lattice.
Additionally we consider the squared curvature defined by
\vspace{-1mm}
\begin{equation}
R^2=\sum_i\frac{\delta_i^2}{A_i}~.
\end{equation}
Furthermore the Liouville mode is of
special interest because it represents the only degree of freedom of pure
2d gravity. The discrete analogue $\phi_i$ of the continuum Liouville field
$\varphi(x)=\ln\sqrt{g(x)}$ is defined by
\begin{equation}
\phi_i = \ln A_i~,~~~A_i=\frac{1}{3}\sum_{t\supset i}A_t~,
\end{equation}
where $A_i$ is the area element of site $i$.
The Liouville susceptibility is then defined by
\begin{equation}
\label{chidef}
\chi_\phi = \langle A \rangle [\langle \phi^2 \rangle
                             - \langle \phi \rangle^2] ~,
\end {equation}
where $\phi = \frac{1}{A}\sum_i \phi_i$ and $A$ is the total area \cite{ham}.
>From Ref. \cite{ham} it is known that for fixed total area $A$ and the scale
invariant measure $dq/q$ the
susceptibility scales according to
\begin{equation} \label{lnchi}
\ln\chi_{\phi}(L) \stackrel{L\to\infty}{\sim} c +
(2 - \eta_{\phi}) \ln L ~,
\end{equation}
with $L=\sqrt{A}$ and the Liouville field critical exponent $\eta_{\phi}=0$.

One can easily derive a scaling relation for the SRC from the partition
function (\ref{CRM}) \cite{ac}. Rescaling all quadratic link lengths of
Eq.~(\ref{CRM}) by a factor $\zeta$, i.e. $q'=\zeta^{-1}q$, yields
\begin{equation}
  \label{scaling}
  Z = \left[ \prod _l \int_0^\infty \zeta^{1-m}
dq{'}_lq{'}_l^{-m} \right] {\cal F}_\eta (\{q{'}_l\})
e^{- \lambda \sum _i \zeta A_i' }~.
\end{equation}
Since this is only a change of the integration variables, $Z$ cannot depend
on $\zeta$. Hence we have
\begin{equation}
  \label{scale_eq}
  \frac{d \ln Z}{d \zeta} |_{\zeta=1} = 0 = N_1 (1-m)
- \lambda \langle \sum _i A_i \rangle
\end{equation}
and find that the expectation value of the area $A$ is fixed to
\begin{equation} \label{area}
  \langle A\rangle =N_1\frac{1-m}{\lambda}=
   N_1\left(\frac{2c_1}{M} +\frac{1}{\lambda}\right)~,
\end{equation}
which is a useful identity to check the correctness of the simulation results.

For positive $\lambda$ the scaling relation makes always sense, and we expect
the partition function (\ref{CRM}) to behave well. However, there is also a
range for negative values of \hbox{$\lambda<\lambda_{c,scal} \equiv -M/2c_1$},
where the area expectation value remains positive. For our choice of
$\epsilon = 0.5$
one obtains $M = 0.5493\dots$ and $c_1 = 0.0788\dots$, such that 
this value equates to
$\lambda_{c,scal}=-3.4817$.
For values of \hbox{$\lambda_{c,scal}<\lambda<0$} the area expectation value
is negative and hence the partition function (\ref{CRM}) is ill-defined.

\section{Simulation details and results} \label{sect4}
We studied the partition functions (\ref{CRM}) and (\ref{Z}) on toroidal
lattices with $N_0 = 16 \times 16 = 256$ sites, resulting in
$N_1 = 768$
and $N_2 = 512$.  The measure was chosen as a function of $\lambda$
according to $m=-\frac{2\lambda c_1}{M} = \lambda/\lambda_{c,scal}$
with $\epsilon = 0.5$, i.e., $m \approx -0.3 \lambda$.  
In the SRC
formulation we used a standard Metropolis algorithm to update every link
during a lattice sweep. For each $\lambda$ in the range of $1-30$,
starting from an initial configuration with equilateral triangles, we first
thermalized the system and then collected 200k measurements taken every
$50^{th}$ MC sweep.  Error bars were computed by the standard Jackknife
method on the basis of 10 blocks. The integrated autocorrelation time
$\tau_A$ for the area was in the range of unity, whereas the integrated
autocorrelation time $\tau_{R^2}$ of $R^2$ was about $100 - 500$, and
$\tau_{\phi}$ of $\phi$ was about $100 - 1000$, depending on the parameter
$\lambda$.

For the update of the spins in the $Z_2$RM formulation we also used
a Metropolis algorithm. Here we performed 100k MC update sweeps measuring
every $10^{th}$ sweep after an initial equilibration period. The integrated
autocorrelation times were typically of the order one.

\begin{figure}[p]
\centerline{\hbox{
\psfig{figure=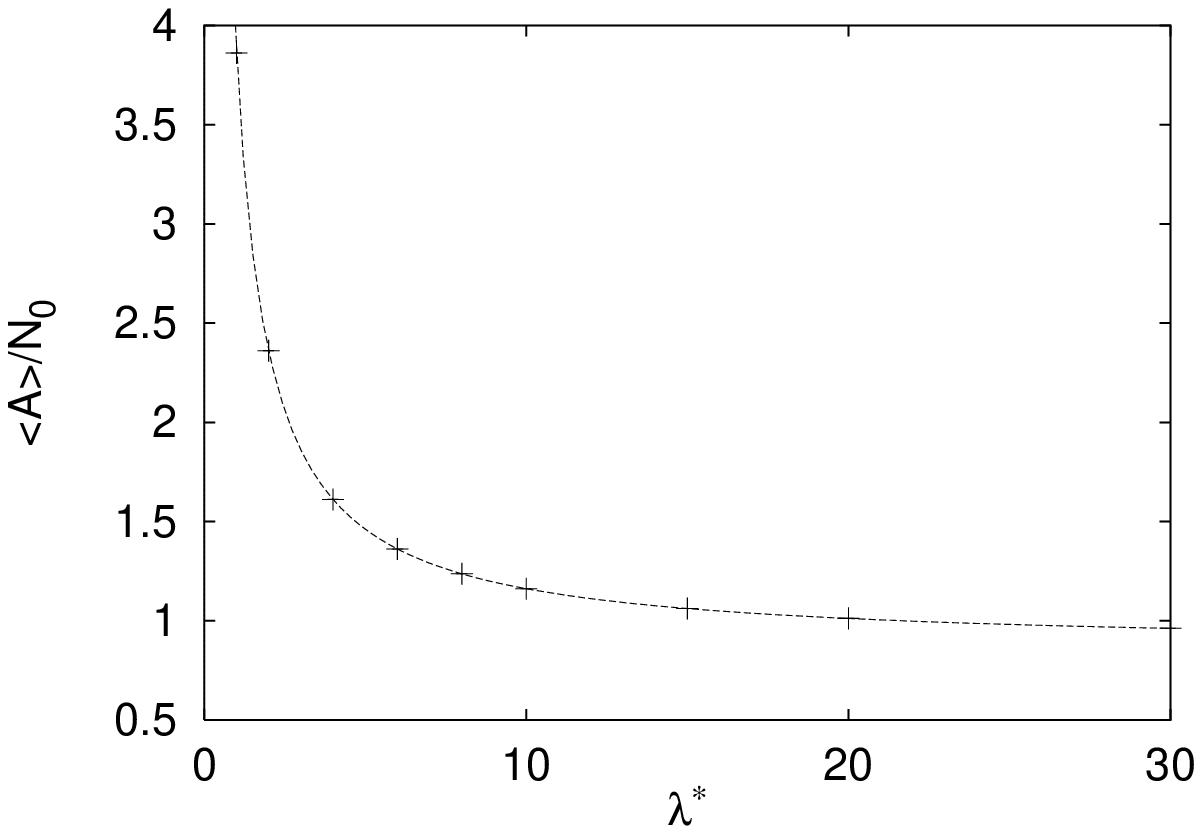,height=6cm,width=8cm}
\psfig{figure=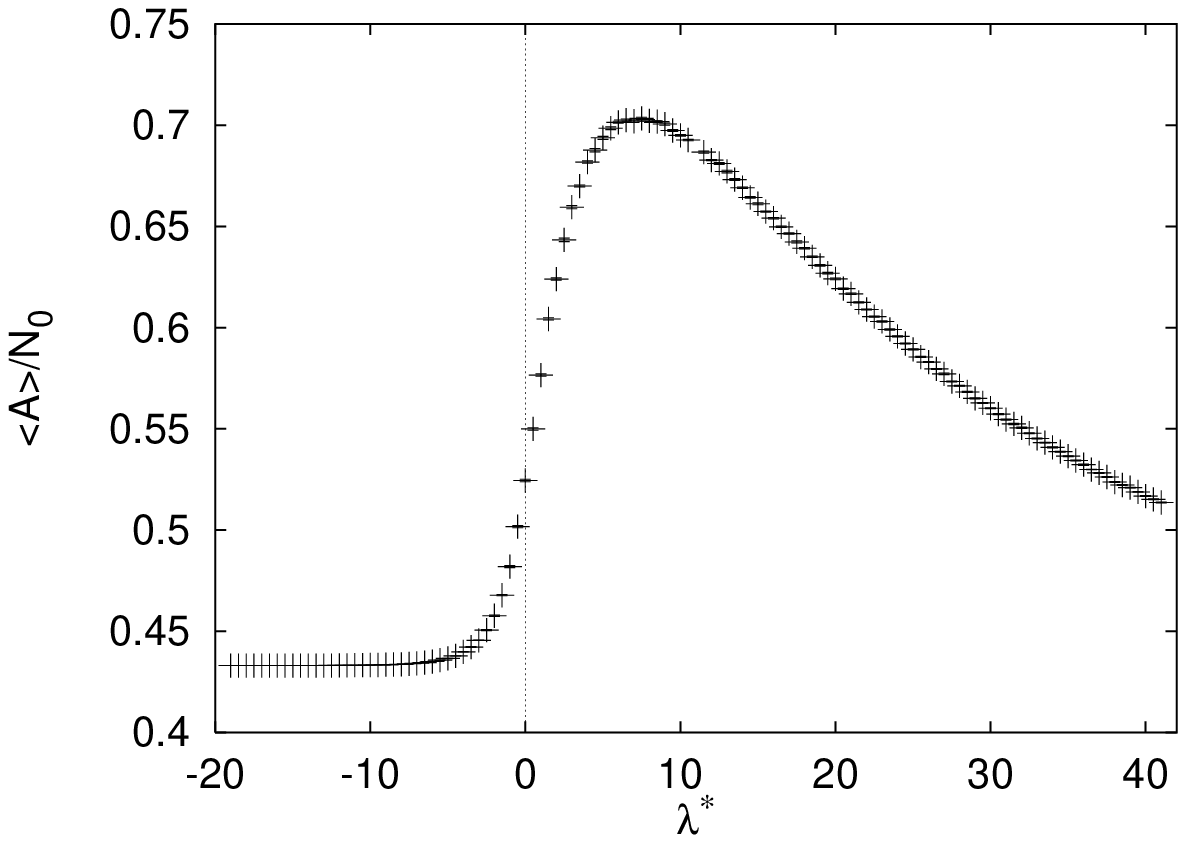,height=6cm,width=8cm}
}}
\centerline{\hbox{
\psfig{figure=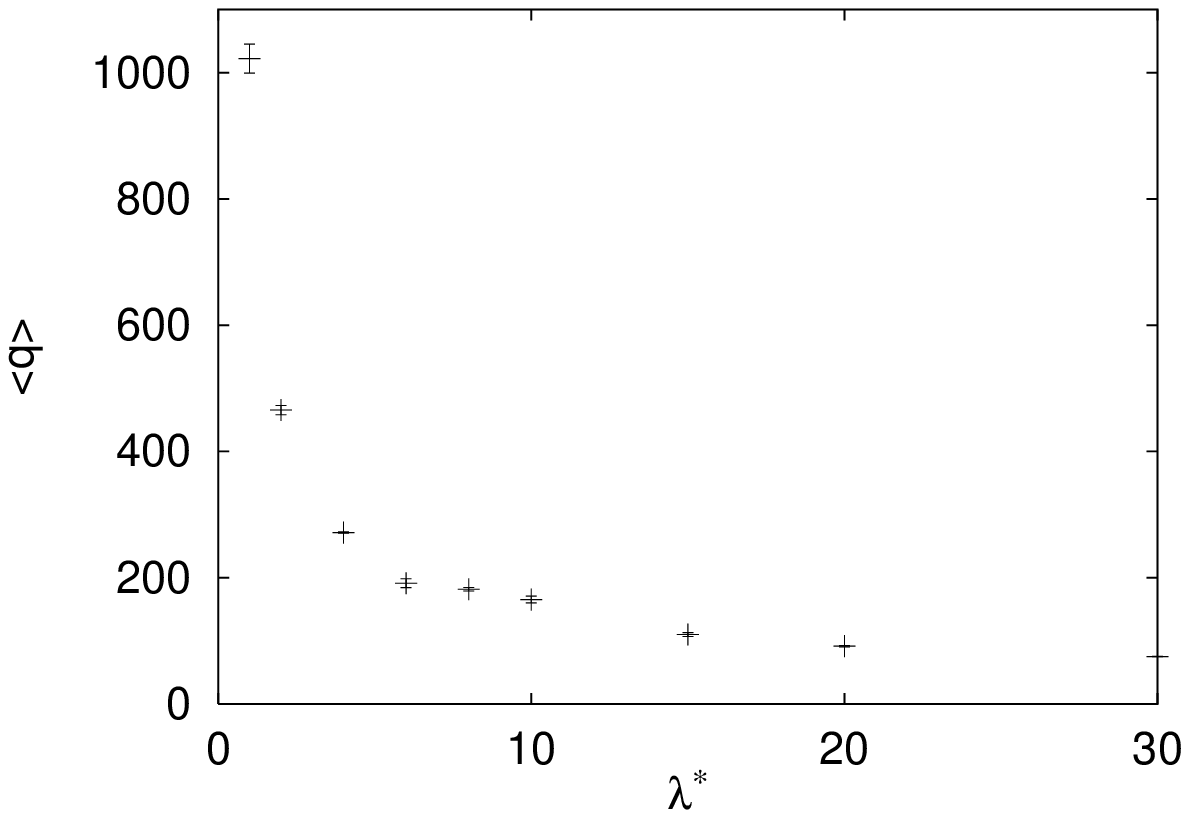,height=6cm,width=8cm}
\psfig{figure=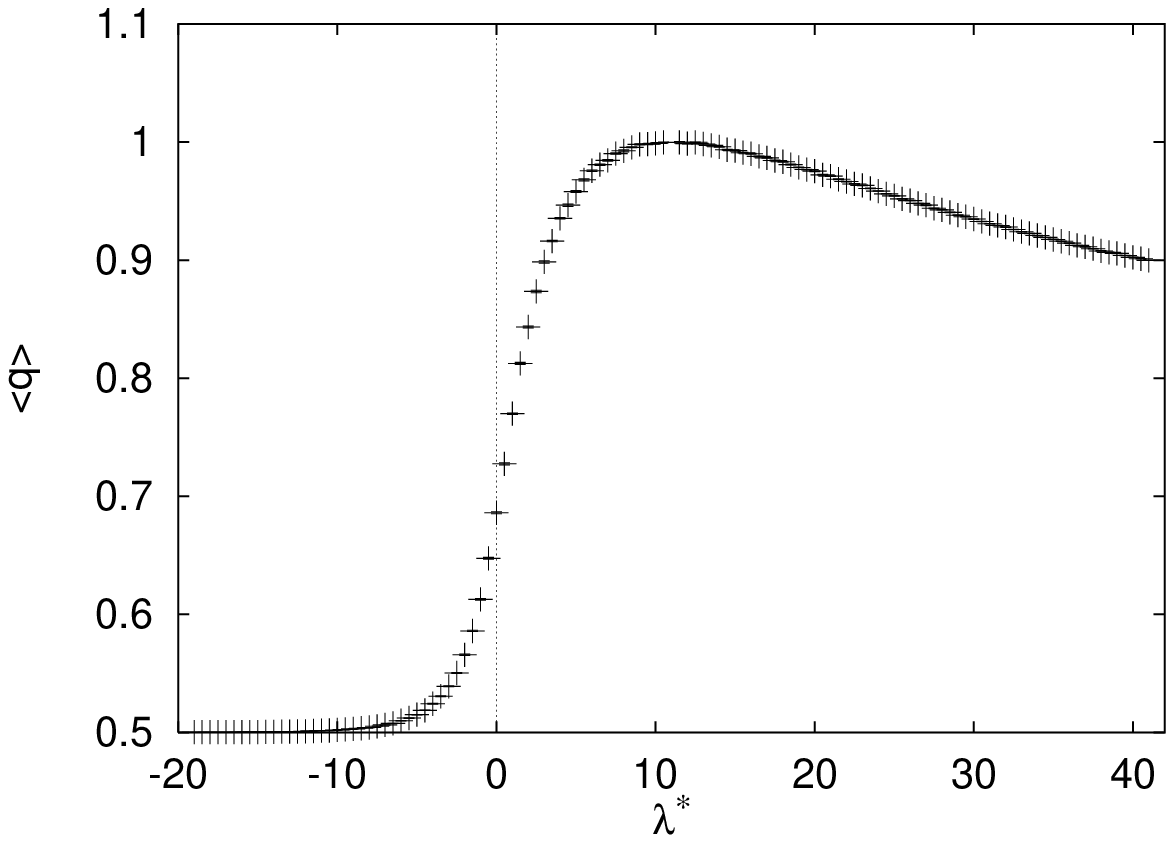,height=6cm,width=8cm}
}}
\centerline{\hbox{
\psfig{figure=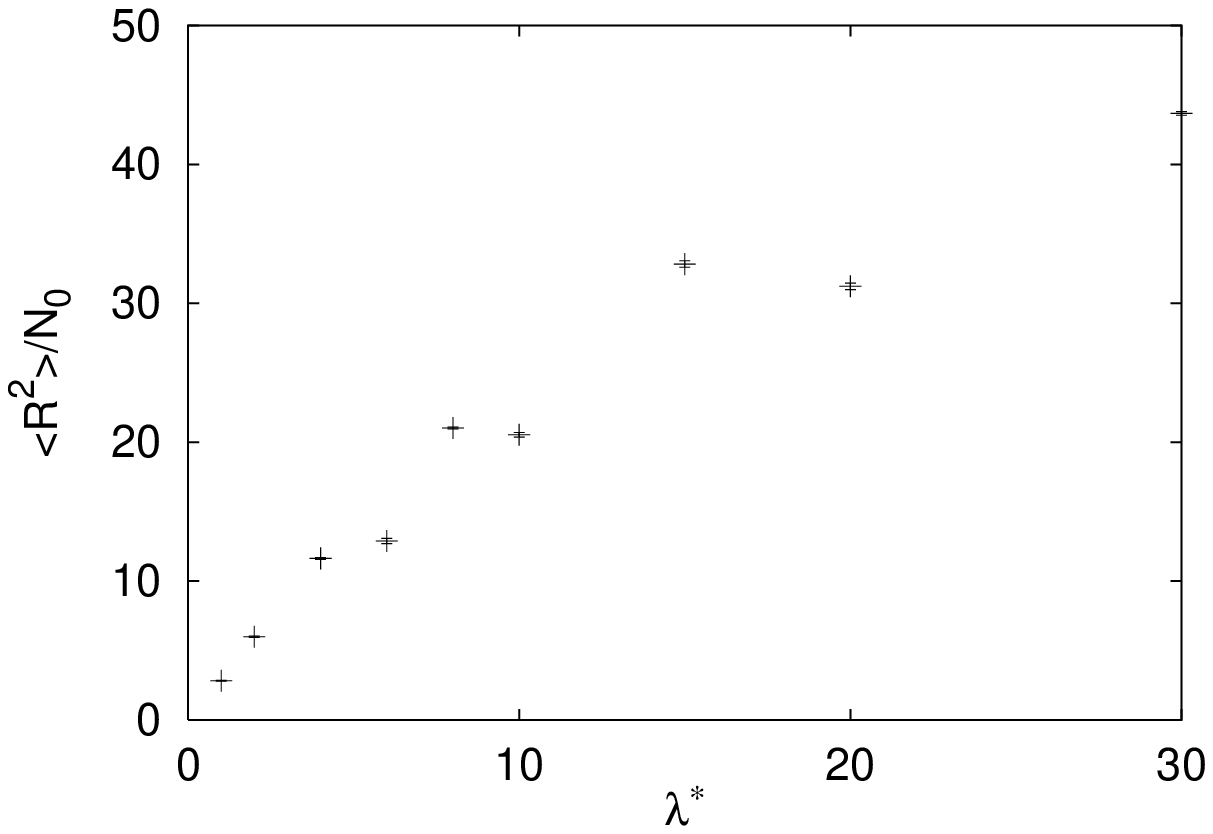,height=6cm,width=8cm}
\psfig{figure=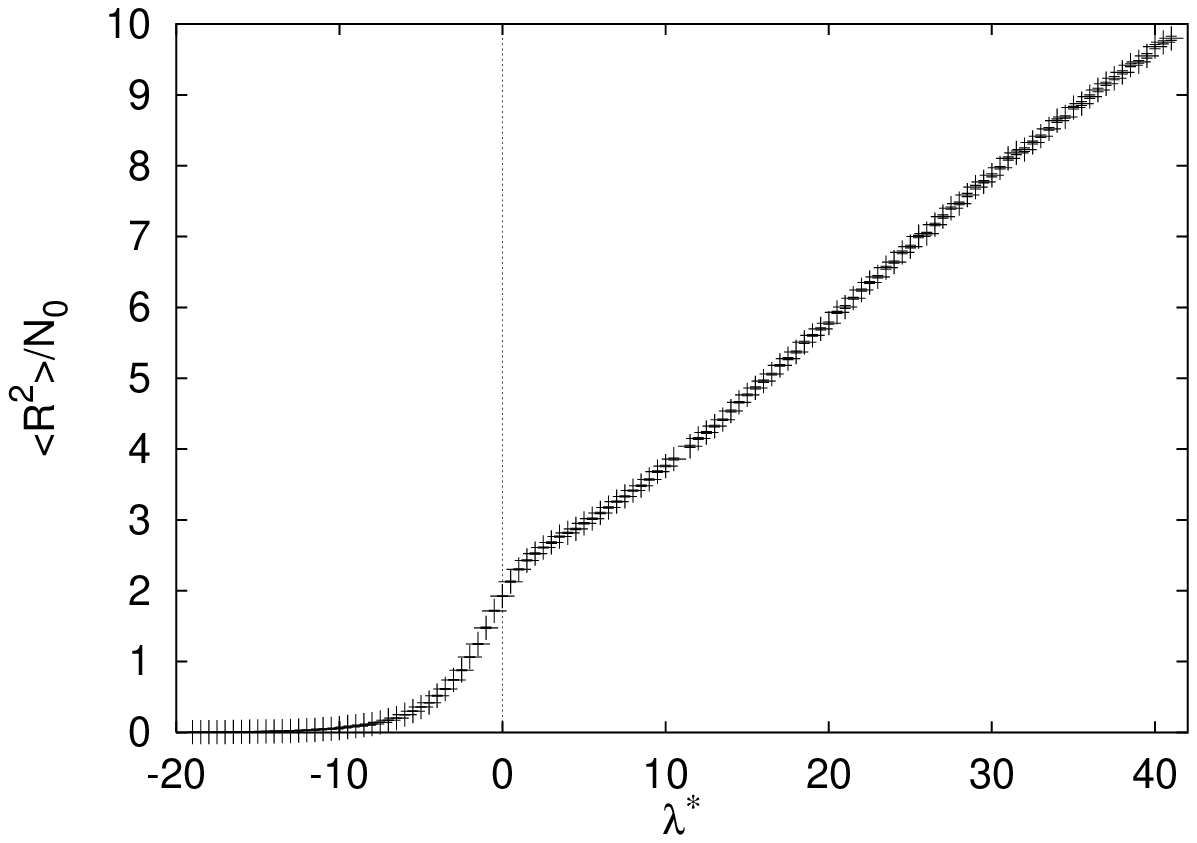,height=6cm,width=8cm}
}}
\vspace{3mm}
\caption{\label{obs1}Expectation values of the area $A$, the average squared
link length $q$, and the squared curvature $R^2$ as a function of the distance
to the critical cosmological constant $\lambda^*$ for the Standard Regge
Calculus (left plots) and the $Z_2$-Regge Model (right plots). $N_0$ is the
total number of vertices.}
\end{figure}

\begin{figure}[p]
\centerline{\hbox{
\psfig{figure=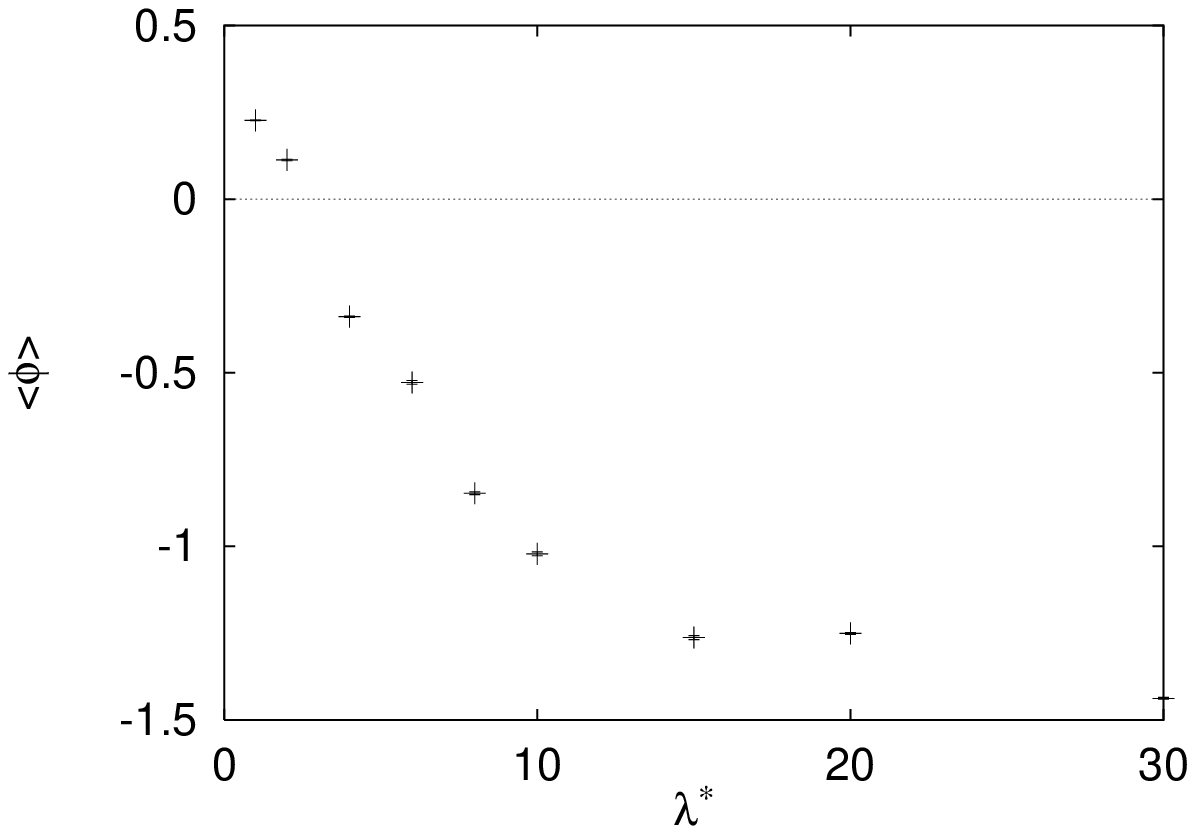,height=6cm,width=8cm}
\psfig{figure=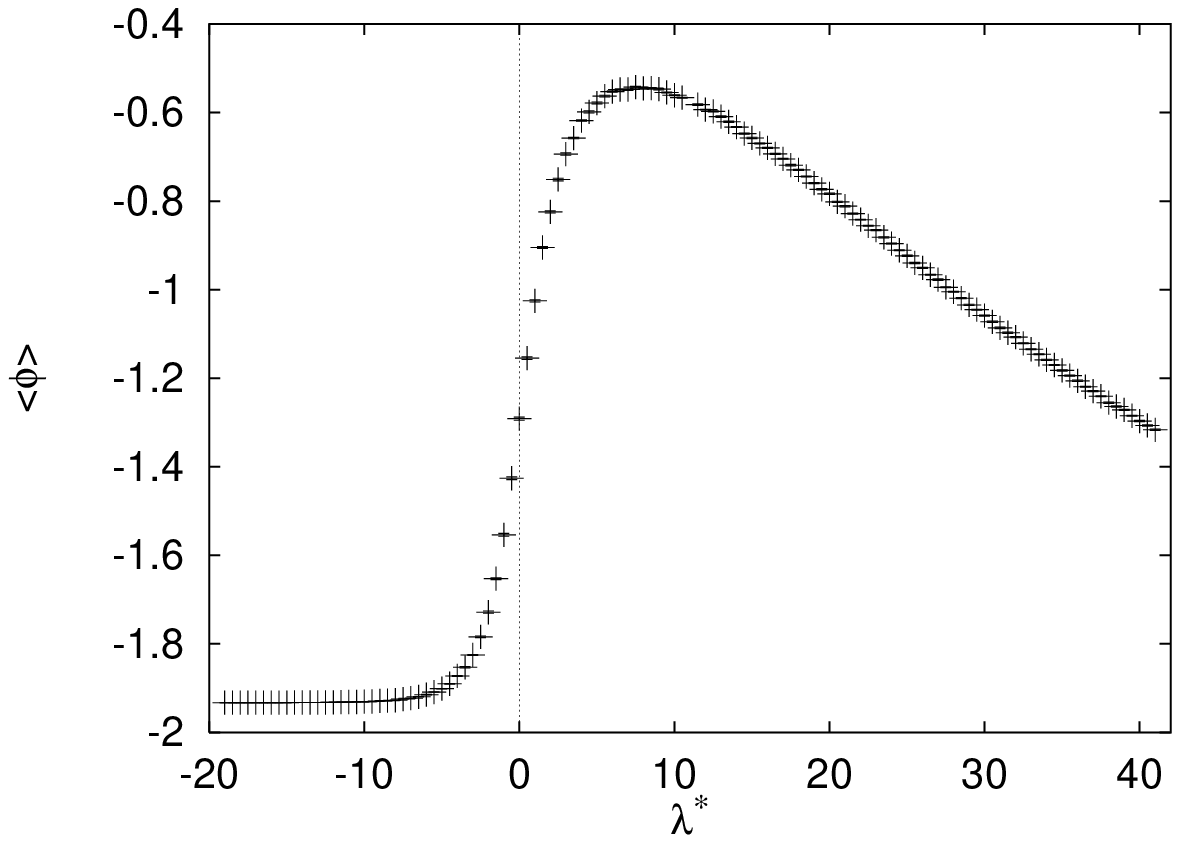,height=6cm,width=8cm}
}}
\centerline{\hbox{
\psfig{figure=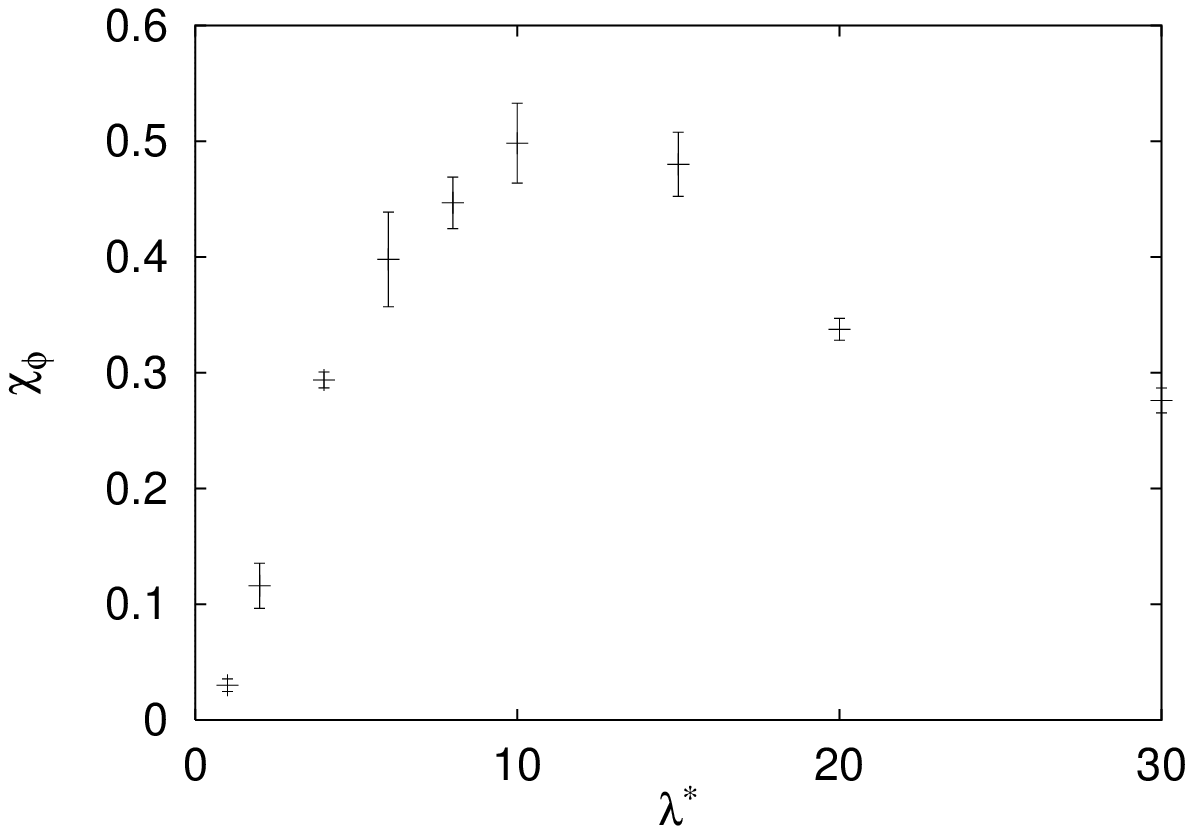,height=6cm,width=8cm}
\psfig{figure=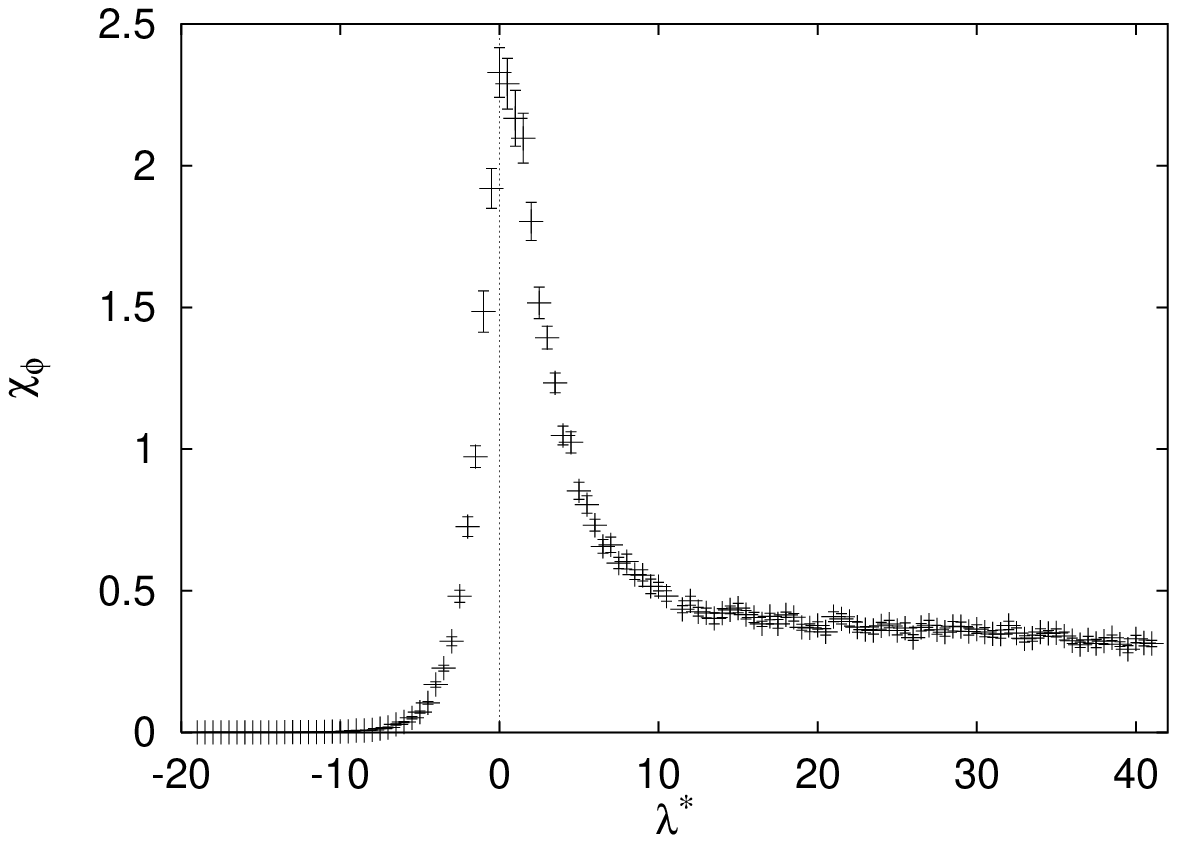,height=6cm,width=8cm}
}}
\vspace{3mm}
\caption{\label{obs2}Expectation values of the Liouville field $\phi$ and the
Liouville field susceptibility $\chi_\phi$ as a function of $\lambda^*$ for
the Standard Regge Calculus (left plots) and the $Z_2$-Regge Model (right
plots).}
\end{figure}

In our first test runs of SRC with positive $\lambda$ we noticed that our
measured area expectation value was not in agreement with the scaling
relation (\ref{area}), and that the mismatch was growing when we increased
$\lambda$. A way to cure this problem was to implement a new global MC
move, which we termed the {\em breathing move\/}. It simply consists of
rescaling
all quadratic link lengths $q$ by a factor $\zeta$, which is the same as
scaling the total area by the factor $\zeta$, where $\zeta$ can be smaller
or larger than one. Again, the usual Metropolis criterion was used to
accept or reject this proposal, and the extent of $\zeta$ was adjusted to
yield an acceptance rate of roughly one half.  We allowed for ten global
breathing moves followed by a complete sweep through the lattice with
normal link updates. Applying these two moves the area expectation value
decreases with increasing $\lambda$ in perfect agreement with the scaling
relation (\ref{area}), see the upper left plot in Fig.~\ref{obs1}.
For sake of better comparison observables are plotted in Fig.~\ref{obs1}
and \ref{obs2} relative to the critical cosmological constant,
$\lambda^*=\lambda-\lambda_c$ ($=\lambda$ for SRC).

With the above global scaling one can also show that the area expectation
value, which follows from the scaling relation for
$\lambda < \lambda_{c,scal}$, is an unstable equilibrium value. This can
be seen by considering the ratio of the Boltzmann factors $B$ before and
after a proposed global move, where we set
$q^{\rm new} = \zeta q^{\rm old}$ and  $A^{\rm new} = \zeta A^{\rm old}$,
with $\zeta = 1 + \kappa$ and $m = 1 - \alpha$. Using these abbreviations
we get
\begin{eqnarray} \label{eq:boltzman}
  \frac{B^{\rm new}}{B^{\rm old}} &=& \frac {\exp (-\lambda\zeta A^{\rm old} +
  \alpha \sum \ln \zeta q^{\rm old})}{\exp (-\lambda A^{\rm old} + \alpha \sum
  \ln q^{\rm old})} \nonumber\\
 &=& \exp ( - \lambda \kappa A^{\rm old} + \alpha \sum \ln (1 + \kappa))
  \\
 &\approx& \exp ( -\kappa \{\lambda A^{\rm old} - \alpha N_1 \} -
  \frac{\alpha N_1}{2} \kappa^2 + \dots )~. \nonumber
\end{eqnarray}
The moves are always accepted in the Monte Carlo simulations if
$\frac{B^{\rm new}}{B^{\rm old}} > 1$. The first term in the argument of the
exponential is linear in $\kappa$ and constitutes
essentially
the scaling
relation (\ref{area}). The second term quadratic in $\kappa$ is {\it always
positive} for negative $\lambda < \lambda_{c,scal}$, i.e., for negative
$\alpha$.
Therefore this term will initially drive the system away from its equilibrium
value. In the next steps, the linear term amplifies the
instability and, 
depending on the arbitrary initial sign of $\kappa$, the area will tend to
zero or infinity.

However, even with the refined update scheme we observed that the SRC system
thermalizes extremely slowly for very small $\lambda$ and therefore display
only statistically reliable data points for \hbox{$\lambda\ge 1$} in the
plots on the l.h.s.\ of Figs.~\ref{obs1} and \ref{obs2}. One also expects
that expectation values of the average squared link length $q$ will decrease
and that expectation values of the squared curvature will increase with
growing $\lambda$ which is indeed confirmed by the simulation results
displayed in Fig.~\ref{obs1}. The Liouville field $\langle \phi \rangle$ and
the associated susceptibility (\ref{chidef}) are shown on the l.h.s.\ of
Fig.~\ref{obs2}. The peak in the Liouville susceptibility can be expected to
move towards $\lambda = 0$ for $L \rightarrow \infty$ due to the diverging
area at this point.

\begin{figure}
\centerline{\hbox{
\psfig{figure=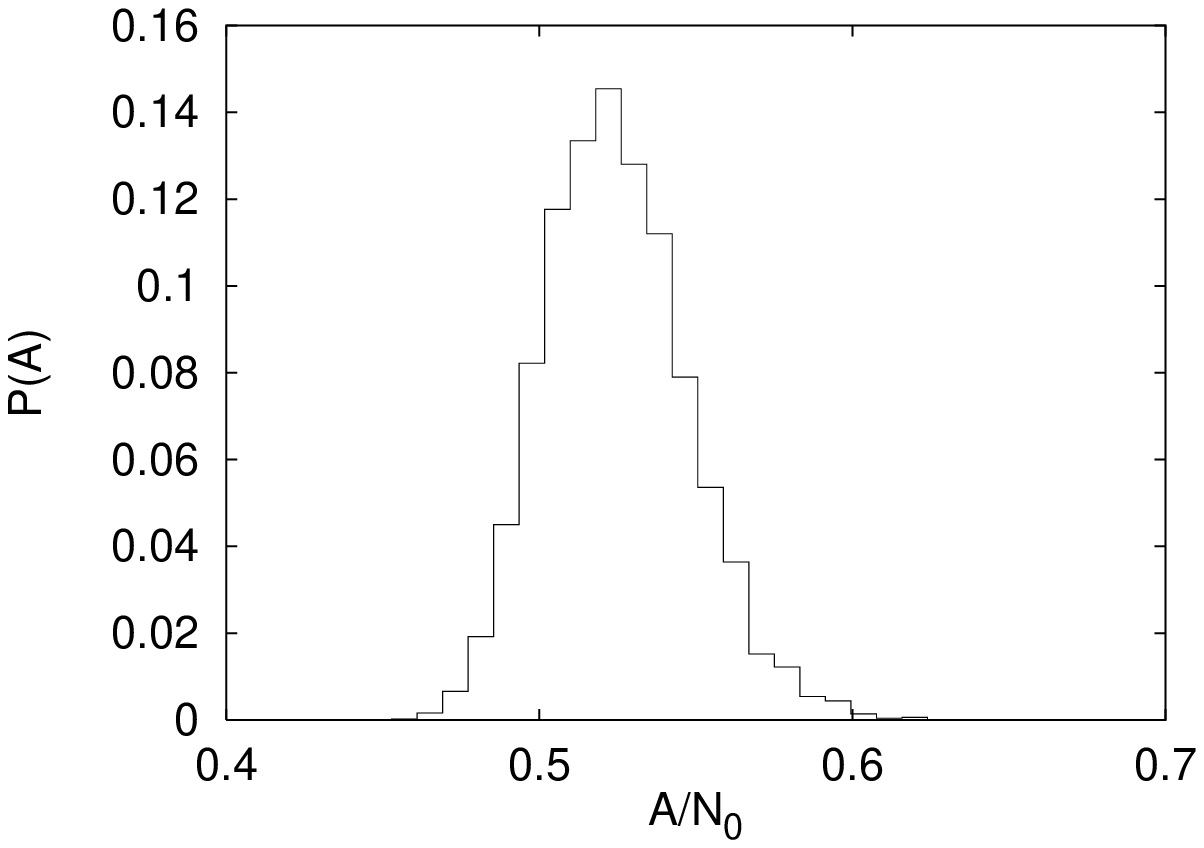,height=6cm,width=8cm}
\psfig{figure=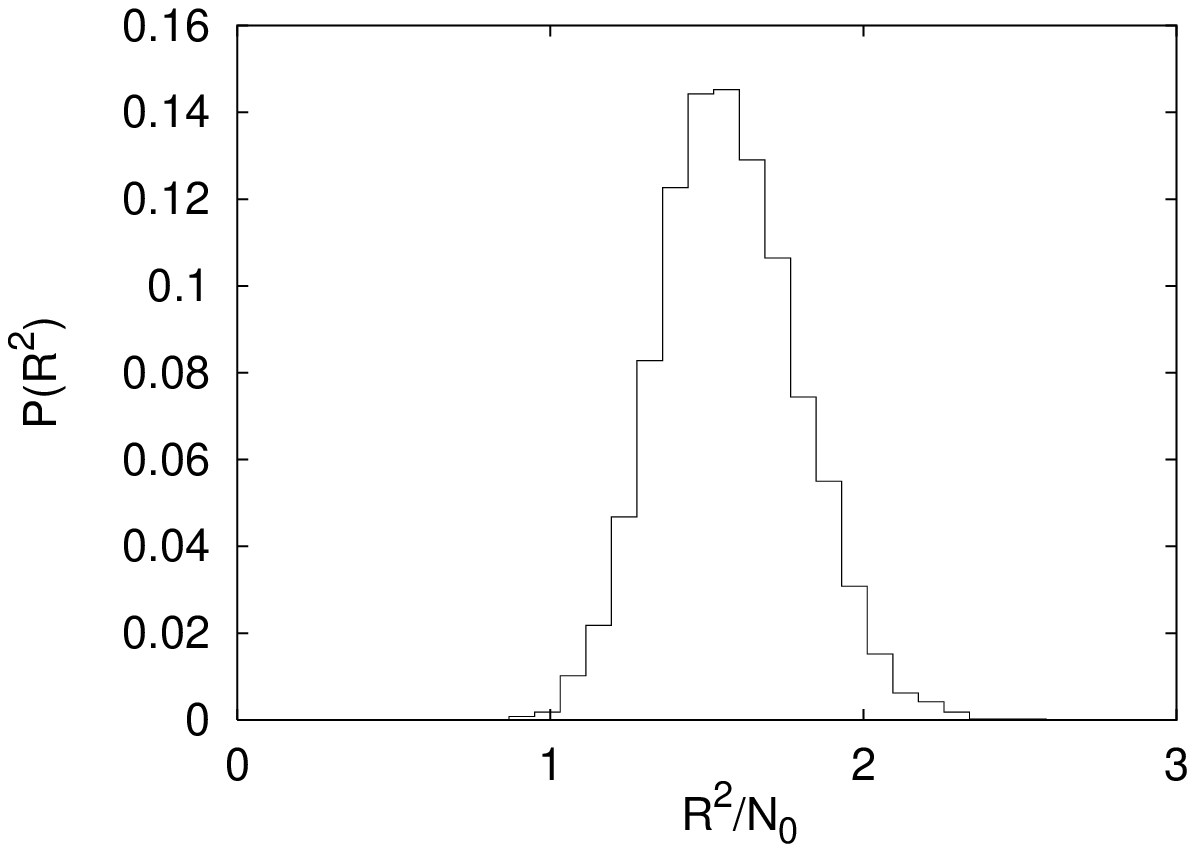,height=6cm,width=8cm}
}}
\vspace{3mm}
\caption{\label{histo}Histograms of the
total
area and the squared
curvature per vertex at the transition point of the $Z_2$-Regge Model.}
\end{figure}

The corresponding quantities of the $Z_2$RM are shown for comparison
on the r.h.s.\ of Figs.~\ref{obs1} and \ref{obs2}. Since it is a crude
approximation a quantitative agreement with SRC
cannot be expected. We see, however, that the two formulations yield
for most quantities the same qualitative behavior for positive couplings
$\lambda^*$.
Whereas the SRC becomes ill-defined for negative $\lambda^*$, the
$Z_2$RM as an effective spin system is well-defined for all values of the
cosmological constant. In all quantities of the $Z_2$RM
we observe the signature of a crossover or phase transition, which is
particularly pronounced in the Liouville susceptibility. From the
peak location we read off a critical coupling $\lambda_c\approx -11$.
The phase transition in the $Z_2$RM might be viewed as some relic of the
transition from a well- to an ill-defined regime of SRC.
To get some idea on the nature of the phase transition we looked at
the histograms
of the
total
area and the squared curvature in Fig.~\ref{histo}. They show
a single-peak structure at $\lambda_c$ which would hint at a continuous
transition.

\begin{figure}
\centerline{\hbox{
\psfig{figure=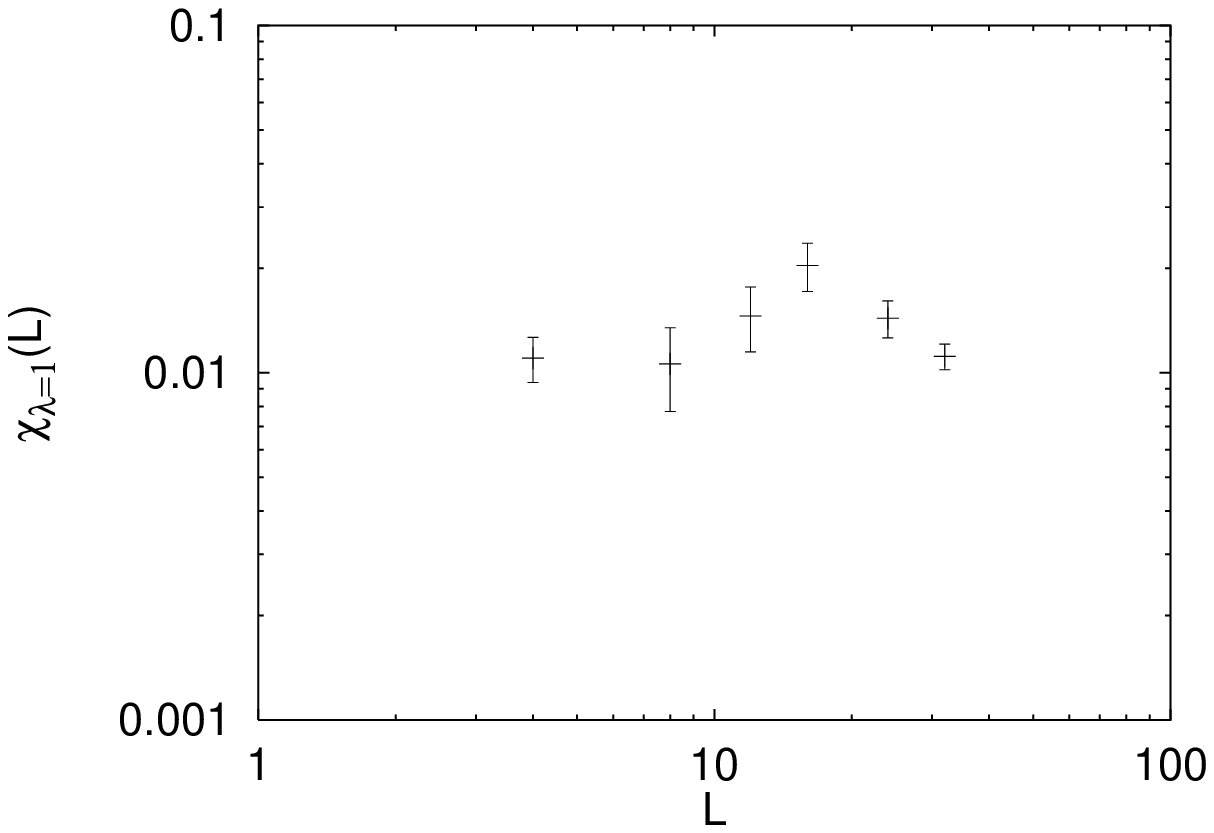,height=6cm,width=8cm}
\psfig{figure=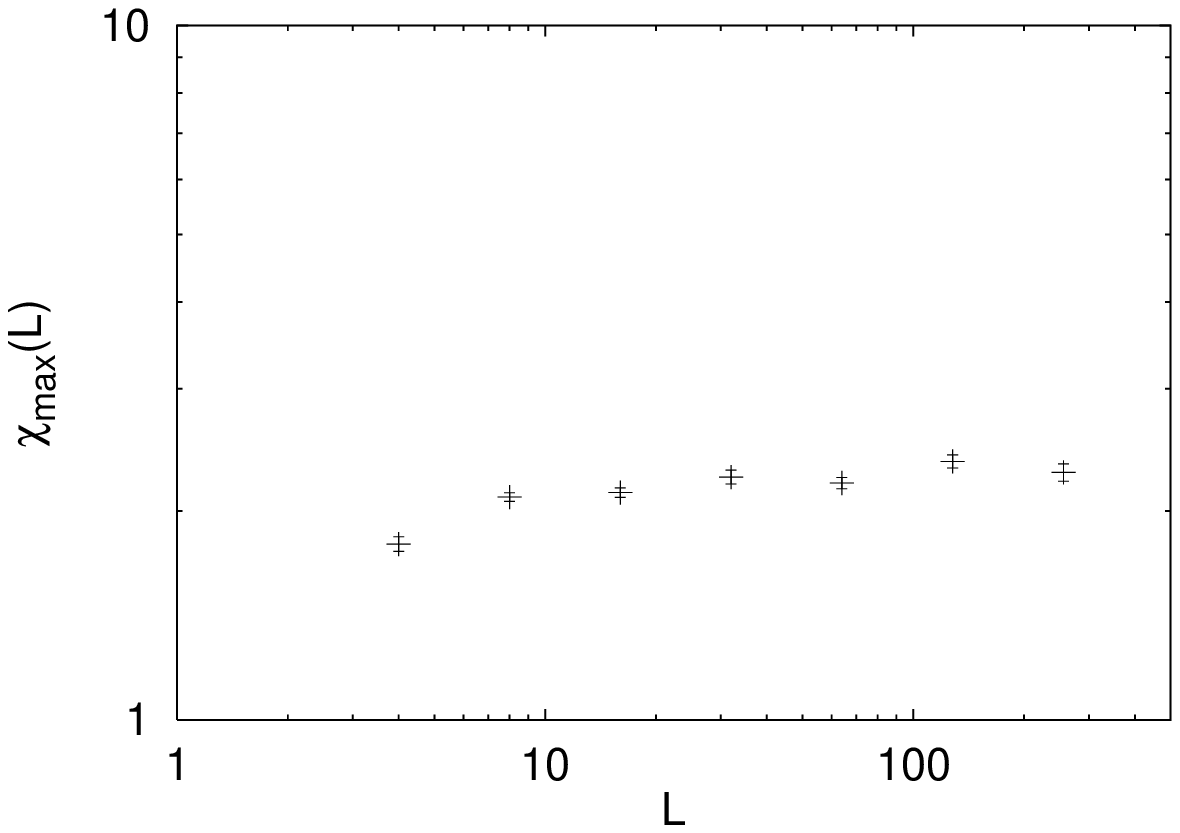,height=6cm,width=8cm}
}}
\vspace{3mm}
\caption{\label{chiscal}Scaling of the of the Liouville field susceptibility
$\chi_\phi$ as a function of the lattice size $L$ for the SRC (left) at
$\lambda=1$ and the $Z_2$RM (right) for max$(\chi_\phi)$.}
\end{figure}

By performing several runs, we further have monitored the scaling of the
Liouville susceptibility maxima
in the $Z_2$RM for lattice sizes up to $L=256$. The double-logarithmic plot
in Fig.~\ref{chiscal} shows that the scaling behavior (\ref{lnchi}) is
governed by a critical exponent $\eta_{\phi}\approx 2$. We have performed a
finite-size scaling analysis of the Liouville susceptibility on lattice sizes
$L = 4$, 8, 12, 16, 24, and 32 also for SRC. Because of the transition
to an ill-defined phase making computations in its vicinity difficult we
employed in all simulations a fixed value $\lambda = 1$ closest to the
critical coupling (see Figs.~\ref{obs1} and \ref{obs2}). The Liouville-field
exponent gives the behavior of the Liouville field in a specific phase,
here the well-defined phase, and hence should be independent of $\lambda$.
After an
equilibration period we performed 20k - 100k measurements of the
Liouville field and computed the associated susceptibility (\ref{chidef}).
Our results presented in Fig.~\ref{chiscal} indeed show that we obtain
the same value of $\eta_\phi\approx 2$, consistent with the results
found in the $Z_2$RM. Furthermore, first results demonstrate scaling
of physically relevant quantities with the number of allowed values for
$\sigma_l$ indicating universality between the $Z_2$RM, SRC and models
``in between'' \cite{L}.

Thus we observe strong disagreement with the Liouville value $\eta_{\phi}=0$,
which has been found for SRC with the scale invariant measure $dq/q$
and fixed area constraint \cite{ham}. There the same
lattice transcription as in the present study was employed
and it seems reasonable to attribute the value
$\eta_{\phi}=0$
to the scale invariance of the measure, whereas
our measure $dq/q^m$ with $m$ given in (\ref{eq:m_vs_M}) explicitly breaks
scale invariance. One therefore might not be surprised to observe a
different value of $\eta_{\phi}$.

\section{Mean-field calculation for the $Z_2$-Regge Model} \label{sect5}
The variational derivation of mean-field theory \cite{ae} starts
with a partition function
\begin{equation}
Z=\sum_{\{\sigma\}}\exp\left[-\beta H(\{\sigma\})\right]~.
\end{equation}
The comparison with equation (\ref{Z}) leads to
\begin{equation}
Z=\sum_{\{\sigma\}} J \exp \left( -\beta \{-\sum_l
\left[ \bar{Q_1}+(\sigma_1+\sigma_2+\sigma_3+\sigma_4)+\bar{Q_3}
(\sigma_1\sigma_2+\sigma_3\sigma_4) \right]\sigma_l \}\right)~,
\end{equation}
with $\beta=-\frac{1}{2}\lambda\,c_2$, $\bar{Q_1} = \frac{4\,c_1}{c_2}
+ \frac{2\,m_1}{\lambda c_2}$ and $\bar{Q_3} = \frac{2\,c_3}{3\,c_2}$.
The spins $\sigma_j, j=1,\dots,4$,
correspond to the squared edge lengths $q_j = 1 + \epsilon \sigma_j$ of the two
triangles which have $q_l$ and hence $\sigma_l$ in common:   

\begin{center}
\unitlength0.5cm
\begin{picture}(6.5,5)(0,0.5)
\thinlines
\put(0,3){\line(1,0){6.5}}
\put(0,2){\line(1,1){4}}
\put(0,3.5){\line(2,-1){6}}
\put(3,6){\line(1,-1){3.5}}
\put(4.5,0){\line(1,2){2}}
\thicklines
\put(1,3){\line(1,0){5}}
\put(1,3){\line(1,1){2.5}}
\put(1,3){\line(2,-1){4}}
\put(3.5,5.5){\line(1,-1){2.5}}
\put(5,1){\line(1,2){1}}
\put(3.4,3.3){$q_l$}
\put(1.6,4.5){$q_1$}
\put(5,4.2){$q_2$}
\put(2.5,1.6){$q_3$}
\put(5.7,2){$q_4$}
\end{picture}

\end{center}

\begin{figure}[p]
\centerline{\hbox{
\psfig{figure=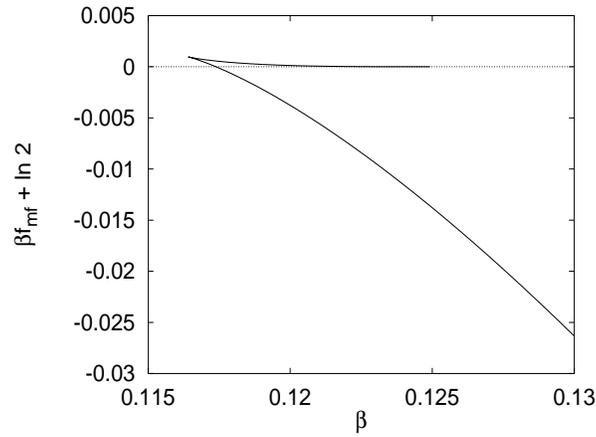,height=6cm,width=8cm}
}}
\vspace{-1mm}
\caption{\label{fig:MF_solution}
Value of $\beta f_{mf}+\ln2$ at the two minima under variation of $\beta$.}
\end{figure}

\begin{figure}[p]
\centerline{\hbox{
\psfig{figure=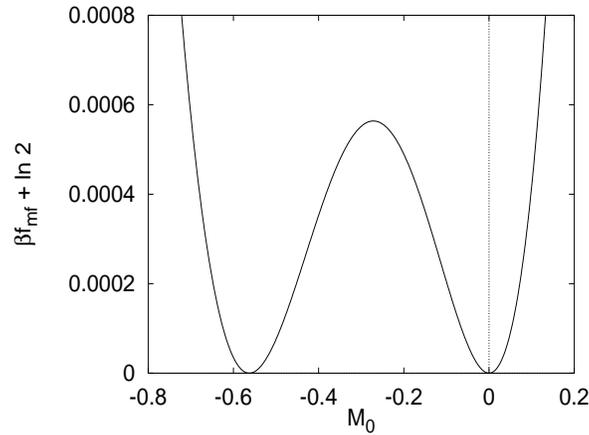,height=6cm,width=8cm}
}}
\vspace{-1mm}
\caption{\label{freeng}
Free energy $\beta f_{mf}+\ln2$ as a function of $M_0$ at the
mean-field transition point.}
\end{figure}

\noindent
Now the Hamiltonian is divided into two parts, $H=H_0+H_1$, where the
choice of $H_0$ is only governed by the requirement that it should be
possible to evaluate the corresponding partition function $Z_0$ analytically.
With the mean-field Ansatz
\begin{equation}
H_0=-\frac{M_0}{\beta}\sum_i\sigma_i
\end{equation}
we get
\begin{equation}
Z_0=\sum_{\{\sigma\}}\exp\left(M_0\sum_i\sigma_i\right) =
\left(e^{M_0}+e^{-M_0}\right)^{N_1} = 2^{N_1}\mbox{cosh}^{N_1}(M_0)
\end{equation}
for the partition function. According to Peierl's inequality we obtain for
\begin{equation}
Z = e^{-\beta F} = Z_0 \langle e^{-\beta
(H-H_0)}\rangle_{_0} \ge Z_0 e^{-\beta\langle H-H_0\rangle_{_0}} =
e^{-\beta F_{m\!f}}
\end{equation}
a lower bound with the mean-field free energy
\begin{equation}
-\beta F_{m\!f} = \ln{Z_0} - \beta \langle H - H_0\rangle_{_0}~.
\end{equation}
The term $\langle \dots\rangle_{_0}$ represents the average with respect to
the system described by $H_0$, i.e.
\begin{eqnarray} \label{03}
\langle H-H_0\rangle_{_0}&=&-\frac{1}{Z_0} \sum_{\{\sigma_i\}}
\sum_l \,\Big[\bar{Q_1}+(\sigma_1+\sigma_2+\sigma_3+\sigma_4)+
\bar{Q_3}(\sigma_1\sigma_2+\sigma_3\sigma_4)- \nonumber \\
&&-\frac{M_0}{\beta}\Big]\,\sigma_l e^{M_0 \sum_i \sigma_i}~.
\end{eqnarray}
Because the thermal average or ``mean'' value of the spin results in
\begin{equation}
\langle \sigma\rangle_{_0} = \frac{e^{M_0}-e^{-M_0}}{e^{M_0}+e^{-M_0}} =
\mbox{tanh}(M_0)~,
\end{equation}
the different terms in (\ref{03}) give
\begin{eqnarray}
\langle \sum_l \sigma_l\rangle_{_0}&=&
N_1 \mbox{tanh}(M_0)~,\nonumber\\
\langle \sum_l [(\sigma_1+\sigma_2+\sigma_3+\sigma_4)\sigma_l]\rangle_{_0}
&=&4 N_1 (\mbox{tanh}(M_0))^2 ~,\\
\langle \sum_l [(\sigma_1\sigma_2+\sigma_3 \sigma_4)\sigma_l]\rangle_{_0}
&=&2 N_1 (\mbox{tanh}(M_0))^3 ~.\nonumber
\end{eqnarray}
Thus we get a $\beta$- and $M_0$-dependent expression for the mean-field
free energy per link,
\begin{eqnarray}
\label{04}
\beta f_{m\!f} \equiv \frac{\beta F_{m\!f}}{N_1}&=&
-\ln{2}-\ln{(\mbox{cosh}(M_0))}-4
\beta (\mbox{tanh}(M_0))^2-2 {\bar Q_3}\beta (\mbox{tanh}(M_0))^3+\nonumber\\
&&+(M_0 - \beta {\bar Q_1}) \mbox{tanh}(M_0)~,
\end{eqnarray}
that has to be minimized ($Z \ge \exp[-\beta F_{m\!f}]$).
Differentiating (\ref{04}) with respect to $M_0$ leads to the
mean-field equation
\begin{equation}
\label{mf_eq}
\mbox{tanh}(M_0)+\frac{3\,{\bar Q_3}}{4} (\mbox{tanh}(M_0))^2 =
\frac{M_0}{8 \beta}-\frac{{\bar Q_1}}{8}
\end{equation}
for the free energy. We restrict ourselves again to a calculation
without external field ${\bar Q_1}$. A trivial solution of (\ref{mf_eq})
is $M_0 = 0$ yielding $\beta f_{m\!f} = -\ln{2}$, which is the stable
solution for small $\beta$ (high ``temperature''). With increasing
$\beta$ a second minimum develops whose free energy is eventually
lower than that of the $M_0 = 0$ solution. This corresponds to the
first-order phase transition expected due to the cubic term in
(\ref{04}). To locate
the transition point
we proceed as follows. For a given $M_0$ value ($< 0$) we can read off
$\beta$ directly from Eq.~(\ref{mf_eq}), without solving any
non-linear equation. By inserting $M_0$ and $\beta$ in (\ref{04})
we obtain immediately the free energy which can then be plotted
versus $\beta$ as in Fig.~\ref{fig:MF_solution}. This curve is
double valued because a given $M_0$ value can correspond to the
minimum (lower branch) or the maximum (upper branch) of
$\beta f_{m\!f}$ at a fixed $\beta$. It should be remarked that for
$\beta > 0.125$ there is another meta-stable solution with $M_0 > 0$
which is not displayed here. With
decreasing $\beta$ the free energy of the non-trivial solution
increases until at $\beta_{c,mf}$ it hits the value of the $M_0=0$
solution. For $\beta < \beta_{c,mf}$ the non-trivial solution first
becomes meta-stable and then disappears completely (at the point
where the minimum- and maximum-branch merge).
In this way it is straightforward to extract
the
transition point as
\begin{equation}
\label{eq:bc}
\beta_{c,mf}\approx 0.1174~~\mbox{corresponding to}~~
\lambda_{c,mf}\approx -8.0~~\mbox{with}~~\epsilon=0.5~.
\end{equation}
At this value the free energy shows a structure with two minima of identical
height (cf. Fig.~\ref{freeng}), which reconfirms that the mean-field
calculation for the $Z_2$-Regge Model predicts a (weak) first-order phase
transition. The value (\ref{eq:bc})
is slightly below the numerical estimate of $\lambda_c \approx -11$; however,
it is a well-known property of mean-field theory to underestimate the
critical coupling. In fact,
the ratio between the mean-field and the numerical value,
$\frac{\lambda_{c,mf}}{\lambda_c}\approx 0.73$, is of the same order of
magnitude as the corresponding quantity for the exactly solvable two
dimensional Ising model on a square lattice where
$(\frac{\lambda_{c,mf}}{\lambda_{c,exact}})_I\approx 0.567$.

\section{Conclusions} \label{sect6}
The aim of this work was to examine whether the $Z_2$RM allowing for two
discrete edge lengths is an appropriate simplification of the SRC with
continuously varying edge lengths in two dimensions. We studied both models
by means of Monte Carlo simulations and found that the simpler $Z_2$RM
qualitatively reproduces the behavior of physical observables like the
Liouville field or the squared curvature for the bare coupling $\lambda>0$.
A finite-size scaling analysis of the Liouville susceptibility
yields for both models the same critical exponent $\eta_{\phi}\approx
2$. This hints at a continuous phase transition which is also necessary to
perform a continuum limit. Concerning universality it is a non-trivial
result, because in the SRC model with scale invariant measure a value of
$\eta_{\phi}=0$ was found.

To obtain further insight, mean-field theory was applied to the $Z_2$RM
indicating a weak first-order phase transition. This would
prevent one to gain important information about the continuum theory.
However, the numerical simulations indicate that fluctuations, which are
neglected in mean-field theory, soften the true nature of the phase
transition to second order. The details of this transition such as
critical exponents etc.\ are still left to be determined.

An interesting question is the influence of allowing for more than two
link lengths and the convergence of the properties of such an extended
$Z_2$RM to those of SRC in two dimensions \cite{L}. With additional degrees
of freedom the situation might resemble the more involved
four-dimensional case where one has to deal with 10 edges per simplex and
the non-trivial Einstein-Hilbert action $\sum_{t\supset i}\delta_tA_t$ with
50 triangles $t$ per vertex $i$ in Eq.~(\ref{Iq}). Thus the action $I(q)$
takes on a large variety of values already for $Z_2$RM and therefore SRC
can be approximated more accurately \cite{4d}.

\section*{Acknowledgments}
A. H. acknowledges support from an Erasmus grant during his stay at
the FU Berlin, where this project was started.
J. R. was supported by Fonds zur F\"orderung der
wissenschaftlichen Forschung under contract P11141-PHY.
W.~J. thanks the Deutsche Forschungsgemeinschaft for a
Heisenberg Fellowship, and also acknowledges partial support by the
German-Israel-Foundation (GIF) under contract No. I-0438-145.07/95.
Parts of the numerical simulations were performed on the North
German Vector Cluster (NVV) under grant bvpf01.


\end{document}